\input harvmac
\input amssym.def
\input amssym.tex
\noblackbox
\newif\ifdraft

\catcode`\@=11
\newif\iffrontpage
\newif\ifxxx
\xxxtrue
%
\newif\ifad
\adtrue
\adfalse

\newif\iffigureexists
\newif\ifepsfloaded
\def\epsfcheck{
\ifdraft
\input epsf\epsfloadedtrue
\else
  \openin 1 epsf
  \ifeof 1 \epsfloadedfalse \else \epsfloadedtrue \fi
  \closein 1
  \ifepsfloaded
    \input epsf
  \else
\immediate\write20{NO EPSF FILE --- FIGURES WILL BE IGNORED}
  \fi
\fi
\def\epsfcheck{}}
\def\checkex#1{
\ifdraft
\figureexistsfalse\immediate%
\write20{Draftmode: figure #1 not included}
\figureexiststrue 
\else\relax
    \ifepsfloaded \openin 1 #1
        \ifeof 1
           \figureexistsfalse
  \immediate\write20{FIGURE FILE #1 NOT FOUND}
        \else \figureexiststrue
        \fi \closein 1
    \else \figureexistsfalse
    \fi
\fi}
\def\missbox#1#2{$\vcenter{\hrule
\hbox{\vrule height#1\kern1.truein
\raise.5truein\hbox{#2} \kern1.truein \vrule} \hrule}$}
\def\lfig#1{
\let\labelflag=#1%
\def\numb@rone{#1}%
\ifx\labelflag\UnDeFiNeD%
{\xdef#1{\the\figno}%
\writedef{#1\leftbracket{\the\figno}}%
\global\advance\figno by1%
}\fi{\hyperref{}{figure}{{\numb@rone}}{Fig.{\numb@rone}}}}
\def\figinsert#1#2#3#4{
\epsfcheck\checkex{#4}%
\def\figsize{#3}%
\let\flag=#1\ifx\flag\UnDeFiNeD
{\xdef#1{\the\figno}%
\writedef{#1\leftbracket{\the\figno}}%
\global\advance\figno by1%
}\fi
\goodbreak\midinsert%
\iffigureexists
\centerline{\epsfysize\figsize\epsfbox{#4}}%
\else%
\vskip.05truein
  \ifepsfloaded
  \ifdraft
  \centerline{\missbox\figsize{Draftmode: #4 not included}}%
  \else
  \centerline{\missbox\figsize{#4 not found}}
  \fi
  \else
  \centerline{\missbox\figsize{epsf.tex not found}}
  \fi
\vskip.05truein
\fi%
{\smallskip%
\leftskip 4pc \rightskip 4pc%
\noindent\ninepoint\sl \baselineskip=11pt%
\smallskip}\bigskip\endinsert%
}

\parindent0pt

\def\k{\kappa}
\def\a{\alpha}
\def\b{\beta}
\def\g{\gamma}

\def\s{\sigma}

\def\p{\partial}

\def\zb{\bar z}

\def\gz{\mathrel{\mathop g^{\scriptscriptstyle{(0)}}}{}\!} 
\def\go{\mathrel{\mathop g^{\scriptscriptstyle{(1)}}}{}\!} 
\def\gt{\mathrel{\mathop g^{\scriptscriptstyle{(2)}}}{}\!} 
\def\gn{\mathrel{\mathop g^{\scriptscriptstyle{(n)}}}{}\!} 
\def\Rz{\mathrel{\mathop R^{\scriptscriptstyle{(0)}}}{}\!} 
\def\Gz{\mathrel{\mathop \Gamma^{\scriptscriptstyle{(0)}}}{}\!} 

\def\{{\lbrace}
\def\}{\rbrace}

\def\R{{\Bbb R}}

\def\A{{\cal A}}
\def\a{\alpha}
\def\b{\beta}

\def\s{\sigma}
\def\g{{g}}

\def\p{\partial}

\def\gz{\mathrel{\mathop g^{\scriptscriptstyle{(0)}}}{}\!\!} 
\def\go{\mathrel{\mathop g^{\scriptscriptstyle{(1)}}}{}\!\!} 
\def\gt{\mathrel{\mathop g^{\scriptscriptstyle{(2)}}}{}\!\!} 
\def\Rz{\mathrel{\mathop R^{\scriptscriptstyle{(0)}}}{}\!\!} 
\def\Xz{\mathrel{\mathop X^{\scriptscriptstyle{(0)}}}{}\!\!} 
\def\Xo{\mathrel{\mathop X^{\scriptscriptstyle{(1)}}}{}\!\!} 
\def\Xt{\mathrel{\mathop X^{\scriptscriptstyle{(2)}}}{}\!\!} 
\def\Xn{\mathrel{\mathop X^{\scriptscriptstyle{(n)}}}{}\!\!} 
\def\hz{\mathrel{\mathop h^{\scriptscriptstyle{(0)}}}{}\!\!} 
\def\ho{\mathrel{\mathop h^{\scriptscriptstyle{(1)}}}{}\!\!} 

\def\sss#1{\scriptscriptstyle{#1}}

\def\Box#1{\mathop{\mkern0.5\thinmuskip
           \vbox{\hrule\hbox{\vrule\hskip#1\vrule height#1 width 0pt\vrule}
           \hrule}\mkern0.5\thinmuskip}}
\def\boxx{\displaystyle{\Box{7pt}}}

\vbadness=10000

\def\frac#1#2{{\scriptstyle{#1}\over\scriptstyle{#2}}}

\def\abstract#1{
\vskip.5in\vfil\centerline
{\bf Abstract}\penalty1000
{{\smallskip\ifx\answ\bigans\leftskip 2pc \rightskip 2pc
\else\leftskip 5pc \rightskip 5pc\fi
\noindent\abstractfont \baselineskip=12pt
{#1} \smallskip}}
\penalty-1000}


\lref\EE{
Some early references on the entanglement entropy are\hfill\break
L.~Bombelli, R.~K.~Koul, J.~H.~Lee and R.~D.~Sorkin,
  ``A Quantum Source of Entropy for Black Holes,''
  Phys.\ Rev.\  D {\bf 34}, 373 (1986); \hfill\break
M.~Srednicki,
  ``Entropy and area,''
  Phys.\ Rev.\ Lett.\  {\bf 71}, 666 (1993)
  [arXiv:hep-th/9303048];\hfill\break
C.~Holzhey, F.~Larsen and F.~Wilczek,
  ``Geometric and renormalized entropy in conformal field theory,''
  Nucl.\ Phys.\  B {\bf 424}, 443 (1994)
  [arXiv:hep-th/9403108].
}

\lref\Fursaev{
D.~V.~Fursaev,
  ``Proof of the holographic formula for entanglement entropy,''
  JHEP {\bf 0609}, 018 (2006)
  [arXiv:hep-th/0606184].
}

\lref\KKM{
  I.~R.~Klebanov, D.~Kutasov and A.~Murugan,
  ``Entanglement as a Probe of Confinement,''
  arXiv:0709.2140 [hep-th].
}
\lref\NT{
  T.~Nishioka and T.~Takayanagi,
  ``AdS bubbles, entropy and closed string tachyons,''
  JHEP {\bf 0701}, 090 (2007)
  [arXiv:hep-th/0611035].}

\lref\Cardy{
P.~Calabrese and J.~L.~Cardy,
  ``Entanglement entropy and quantum field theory,''
  J.\ Stat.\ Mech.\  {\bf 0406}, P002 (2004)
  [arXiv:hep-th/0405152];
  ``Entanglement entropy and quantum field theory: A non-technical introduction,''
  Int.\ J.\ Quant.\ Inf.\  {\bf 4}, 429 (2006)
  [arXiv:quant-ph/0505193].
}

\lref\RT{
S.~Ryu and T.~Takayanagi,
  ``Holographic derivation of entanglement entropy from AdS/CFT,''
  Phys.\ Rev.\ Lett.\  {\bf 96}, 181602 (2006)
  [arXiv:hep-th/0603001];
  ``Aspects of holographic entanglement entropy,''
  JHEP {\bf 0608}, 045 (2006)
  [arXiv:hep-th/0605073].
}

\lref\ISTY{
C.~Imbimbo, A.~Schwimmer, S.~Theisen and S.~Yankielowicz,
  ``Diffeomorphisms and holographic anomalies,''
  Class.\ Quant.\ Grav.\  {\bf 17}, 1129 (2000)
  [arXiv:hep-th/9910267].
}

\lref\WittenI{
E.~Witten,
  ``Anti-de Sitter space and holography,''
  Adv.\ Theor.\ Math.\ Phys.\  {\bf 2}, 253 (1998)
  [arXiv:hep-th/9802150].
}

\lref\FG{
C.~Fefferman and R.~Graham, ``Conformal Invariants,''
Ast\`erisque, hors s\'erie, 1995, p.95.}

\lref\HS{
M.~Henningson and K.~Skenderis,
  ``The holographic Weyl anomaly,''
  JHEP {\bf 9807}, 023 (1998)
  [arXiv:hep-th/9806087].
}

\lref\AR{
A.~D.~Rendall,
  ``Asymptotics of solutions of the Einstein equations with positive
  cosmological constant,''
  Annales Henri Poincare {\bf 5}, 1041 (2004)
  [arXiv:gr-qc/0312020].
}

\lref\GW{
C.~R.~Graham and E.~Witten,
  ``Conformal anomaly of submanifold observables in AdS/CFT correspondence,''
  Nucl.\ Phys.\  B {\bf 546}, 52 (1999)
  [arXiv:hep-th/9901021].
}

 \lref\FS{
D.~V.~Fursaev and S.~N.~Solodukhin,
 ``On the Description of the  Riemannian Geometry in the Presence of
 Conical Defects,'' Phys.\ Rev.\ D{\bf 54}, 2133 (1995)
 [arXiv:hep-th/9501127].
}

 \lref\Boulanger{ N.~Boulanger,
  ``General solutions of the Wess-Zumino consistency condition for the Weyl
  anomalies,''
  JHEP {\bf 0707}, 069 (2007)
  [arXiv:0704.2472 [hep-th]];
  ``Algebraic Classification of Weyl Anomalies in Arbitrary Dimensions,''
  Phys.\ Rev.\ Lett.\  {\bf 98}, 261302 (2007)
  [arXiv:0706.0340 [hep-th]].
}

\lref\HStwo{M.~Henningson and K.~Skenderis,
  ``Weyl anomaly for Wilson surfaces,''
  JHEP {\bf 9906}, 012 (1999)
  [arXiv:hep-th/9905163].
}

\lref\Asnin{
V.~Asnin,
  ``Analyticity Properties of Graham-Witten Anomalies,''
  arXiv:0801.1469 [hep-th].
}

\lref\Gustavsson{
A.~Gustavsson,
  ``Conformal anomaly of Wilson surface observables: A field theoretical
  computation,''
  JHEP {\bf 0407}, 074 (2004)
  [arXiv:hep-th/0404150];
  ``On the Weyl anomaly of Wilson surfaces,''
  JHEP {\bf 0312}, 059 (2003)
  [arXiv:hep-th/0310037].
}

\lref\WZ{
J.~Wess and B.~Zumino,
  ``Consequences of anomalous Ward identities,''
  Phys.\ Lett.\  B {\bf 37}, 95 (1971).
}

\lref\Cappelli{
  A.~Cappelli and A.~Coste,
  ``On the Stress Tensor of Conformal Field Theories in Higher Dimensions,''
  Nucl.\ Phys.\  B {\bf 314}, 707 (1989);
E.~S.~Fradkin and A.~A.~Tseytlin,
  ``Conformal Anomaly In Weyl Theory And Anomaly Free Superconformal
  Theories,''
  Phys.\ Lett.\  B {\bf 134}, 187 (1984).
}

\lref\BFS{
M.~Bianchi, D.~Z.~Freedman and K.~Skenderis,
  ``How to go with an RG flow,''
  JHEP {\bf 0108}, 041 (2001)
  [arXiv:hep-th/0105276];
``Holographic renormalization,''
  Nucl.\ Phys.\  B {\bf 631}, 159 (2002)
  [arXiv:hep-th/0112119].
}

\lref\ST{
A.~Schwimmer and S.~Theisen,
  ``Universal features of holographic anomalies,''
  JHEP {\bf 0310}, 001 (2003)
  [arXiv:hep-th/0309064].
}

\lref\PS{I.~Papadimitriou and K.~Skenderis,
  ``Thermodynamics of asymptotically locally AdS spacetimes,''
  JHEP {\bf 0508}, 004 (2005)
  [arXiv:hep-th/0505190].
}

\lref\AdSreview{
O.~Aharony, S.~S.~Gubser, J.~M.~Maldacena, H.~Ooguri and Y.~Oz,
``Large N field theories, string theory and gravity,''
Phys.\ Rept.\  {\bf 323}, 183 (2000), hep-th/9905111.}


\Title{\vbox{
\rightline{\vbox{\baselineskip12pt}}}}
{Entanglement Entropy, Trace Anomalies and Holography
\footnote{$^{\scriptscriptstyle*}$}{\sevenrm
Partially supported by
the German-Israeli Project cooperation (DIP H52), the Einstein Center
of the Weizmann Institute, the Humboldt Foundation and
by the European Research and Training Networks
`Superstrings' (MRTN-CT-2004-512194) (A.S.) and
`Forces Universe' (MRTN-CT-2004-005104) (S.T.).
}}
\vskip0.3cm
\centerline{A.~Schwimmer$^a$ and S.~Theisen$^b$}
\vskip 0.6cm
\centerline{$^a$ \it Department of Physics of Complex Systems,
Weizmann Institute, Rehovot 76100, Israel}
\vskip.2cm
\centerline{$^b$ \it Max-Planck-Institut f\"ur Gravitationsphysik,
Albert-Einstein-Institut, 14476 Golm, Germany}
\vskip0.0cm

\abstract{The holographic representation of the entanglement entropy of four
dimensional conformal field theories is studied. By generalizing
the replica trick the anomalous terms in the entanglement entropy
are evaluated. The same terms in the
holographic representation are calculated by a method which does
not require the solution of the equations of motion or a cut off.
The two calculations disagree for rather generic geometries. The
reasons for the disagreement are analyzed.}
\Date{\vbox{\hbox{\sl {}}
}}
\goodbreak

\newsec{Introduction}

Entanglement entropy (\EE) was proposed as a powerful tool for
studying in detail the structure of vacua in QFT.

In $d=2$ CFT  universal features of entanglement entropy (EE in the
following) were studied in \Cardy: in particular the
dependence of the EE on the scale of the region defining the EE was
shown to be related to the trace anomaly of the CFT.

These results were generalized to any $d$ in \RT. Following the
$d=2$ example \Cardy\ it was postulated that  the EE in flat space
is given by the usual partition function when the theory is defined
in a metric with a conical singularity supported on a submanifold of
codimension $2$. The submanifold is the boundary between the two
regions, $A$ and $B$, defining the EE. Universal terms in the scale
dependence of the EE were again estimated by relating them to the
trace anomalies (when $d$ is even) of the CFT. A very interesting
proposal was made in \RT\ for a holographic realization of the EE:
if the theory in flat space has a holographic dual $AdS_{d+1}\times
X$ then the EE is realized  by the above bulk gravitational theory
to which one adds a $(d-1)$ Dirac-Nambu-Goto action (DNG action in
in the following) in the $AdS$ background, the coordinates at the
boundary of $AdS$ representing the embedding of the
$(d-2)$-dimensional boundary manifold.

This prescription allows in principle the calculation of all the
observables in the EE at strong coupling including the complete
scale dependence.

The prescription presupposes, as we will discuss in detail, that the
theory in the singular metric can be represented equivalently by a
smooth bulk metric to which an additional term defined on the
singular manifold is added. Arguments for the validity of this
assumption were presented in \Fursaev. Using the aforementioned
holographic representation for massive $d=4$ theories in \KKM\
information about the phase structure of these theories was
obtained; see also \NT\ for an earlier discussion of this issue.

In the present paper we reexamine the validity of the holographic
prescription for the four dimensional EE. We concentrate on the
terms in the $d=4$ effective action representing the trace
anomalies. These terms, being universal, can be controlled in a
general CFT. Following the ``replica trick'' we deduce a singular
metric which represents the EE. We assume that at least for the
terms controlling the trace anomalies one can use the singular
metric in the effective action without further regularization. We
also assume that  the ``replica trick'' expressions can be safely
expanded to first order.

We conclude that for very generic cases (which include e.g. the
situation when the boundary of the region defining the EE is a
two-sphere) one cannot represent the singular metric by an
additional piece localized on the singular submanifold. From a
detailed study of the term in the effective action which generates
the type A (Euler) trace anomaly we find that if the term is split
into a $d=4 $ part in a smooth metric and a $d=2$ part localized on
the singular manifold the $d=2$ part does not have correct
analyticity properties. If the splitting is done after a Weyl
variation is taken (i.e. for the anomaly itself) the $d=2$ part does
not fulfill the Wess-Zumino conditions. Since, as we will discuss in
detail  the holographic realization mentioned above has  anomalous
terms (bulk and Graham-Witten anomalies) with unambiguous
``normal'' analyticity properties it follows that the holographic
prescription is not valid, at least for this particular, universal
piece of the effective action. The analyticity requires that certain
Bianchi identities are satisfied but the identities require
contributions from singular and regular terms, preventing a
consistent ``splitting''. The discrepancy between the CFT and its
proposed holographic realization appears whenever the second
fundamental form of the submanifold is nonvanishing.

One possibility we examine for explaining the discrepancy is to
include the back reaction of the DNG action on the bulk component. It
turns out  that the back reaction does not change the anomalous
terms.

Therefore another option for a holographic realization of the EE is
the usual five dimensional bulk action where at the boundary one
matches to the singular four dimensional metric. This prescription
can be implemented at least for the anomaly calculations since, as
we will discuss in detail, in this case one does not need to solve
the classical equations of motion the anomalies being given by a
direct evaluation of certain boundary terms. The validity of this
straightforward prescription for other terms representing the EE as
well as the origin of nonuniversal contributions to the EE requires
further study.

The holographic prescription containing the DNG action has the
correct analytic structure to represent genuine extended
observables in the CFT as it was shown explicitly in \HStwo,
\Gustavsson\ and \Asnin.

The paper is organized as follows:
In Section 2  we generalize the replica trick of \Cardy\ to
$d=4$ and we deduce an explicit form for the singular metric in the
most general case needed for EE.

In Section 3 we discuss in detail the calculation of trace anomalies
in the holographic set up both for the ones originating in the bulk
and from the DNG piece of the action (``Graham-Witten anomalies'').
We show that the calculation of the anomalies does not use the
solution of the equations of motion reducing to the evaluation of a
total derivative on the boundary. Furthermore this evaluation does
not require the use of any cutoff procedure. Therefore the
analyticity properties of the terms of the effective action
responsible for the anomalies can be generally obtained.

In Section 4 we discuss the structure of the anomalous terms in the
$d=4$ CFT in the singular metric representing the EE. We show that
there is a contradiction between the structure of the type A (Euler)
term and the one expected from the holographic description discussed
in Section 3. We discuss the general mechanism which prevents the
``splitting'' of the  EE problem into a bulk part and an additional
piece formulated on the singular submanifold.

In Section 5 we study the influence of including the back reaction
of the DNG action on the bulk. We show by a general argument that
the back reaction does not change the anomalous terms. The argument
is verified by solving the coupled equations of motion
to the necessary order.

In Section 6 we calculate explicitly universal pieces of the EE for
a sphere  and we show that the results extracted from the CFT
following our procedure disagree with the holographic prescription.
In the same section we discuss open problems related  to terms not
controlled by the trace anomalies.

A new proof of a universal relation for the type A bulk trace
anomaly \ISTY\ using the approach of Section 3 is presented in the
Appendix.

\newsec{EE in $d=4$}
We start by discussing the four dimensional generalization of the
replica trick \Cardy\ we are going to use.

Consider a conformal field $\phi(\vec x,t)$ in ${\R}^{3,1}$. To simplify the notation
we will consider a scalar field but our arguments should be valid
for any spin. Let
$A$ be a region in ${\R}^3$ and $B$ its complement.\foot{For
simplicity one might assume $A$ to be compact and connected.} The
boundary between $A$ and $B$ will be denoted $\p$.
We will start by assuming that the metric of space time is flat
but will later generalize it to a situation where the metric
components can be space dependent such that the EE can still be
defined.

The density
matrix of the system has the path integral representation
\eqn\denmat{
(\rho)_{\phi'(\vec x',0),\phi''(\vec x'',0)}={1\over Z}\int {\cal D}\phi(x)
\prod_x\delta(\phi(\vec x,0^-)-\phi'(\vec x',0))\,
\delta(\phi(\vec x,0^+)-\phi''(\vec x'',0))\,
e^{-S[\phi]}}
The reduced density matrix is obtained by setting $\phi'(\vec x,0)=\phi''(\vec x,0)$ for $x\in B$
and integrating over $\phi$ which are continuous across $B$:
\eqn\rhored{
(\rho_A)_{\phi'(\vec x',0),\phi''(\vec x'',0)}={1\over Z}\int {\cal D}\phi(x)
\prod_{x\in A}\delta(\phi(\vec x,0^-)-\phi'(\vec x',0))\,
\delta(\phi(\vec x,0^+)-\phi''(\vec x'',0))\,
e^{-S[\phi]}}

The entanglement entropy $S_A$ is the von Neumann entropy computed with the reduced
density matrix $\rho_A$
\eqn\SAdef{
S_A=-{\rm Tr}_A\,\rho_A\log\rho_A}
Various properties of the entanglement entropy are collected
in \RT.

The ``standard'' replica trick, by which
\eqn\SAlim{
\rho_A=-\lim_{n\to 1}{\p\over\p n}{\rm Tr}\rho_A^n}
leads to $n$ copies $\phi^{(i)}$ of the field $\phi$ linked by the boundary conditions
\eqn\bcphiA{
\eqalign{
\phi^{(i)}_A(\vec x,t=0^+)\quad &=\quad\phi^{(i+1)}_A(\vec x,t=0^-) \cr
}}
where $\phi_A$ means that $\vec x\in A$ and $n+1\equiv 1$.
In addition in region $B$ the fields are continuous, i.e. single valued at $t=0$:
\eqn\bcphiB{
\eqalign{
\phi^{(i)}_B(\vec x,t=0^+)\quad &=\quad\phi^{(i)}_B(\vec x,t=0^-)\cr
}}
where $\phi_B$ means $\vec x\in B$.

The action appearing in the path integral is
\eqn\actionAB{
S=\sum_{i=1}^nS[\phi_A^{(i)},\phi_B^{(i)}]}
with the gluing specified in \bcphiA,\bcphiB. It is important that in the path integral
there are no singularities: the time derivatives around $t=0$ are for $A$:
\eqn\timeder{
\left({\p\phi\over\p t}\right)^2\to\left({\phi^{(1)}(t+\epsilon)-\phi^{(2)}(t-\epsilon)\over 2\epsilon}\right)^2
\qquad\hbox{etc.}}
Now define a new set of fields $\tilde\phi$:
\eqn\newphi{
\eqalign{
\tilde\phi_A^{(i)}(\vec x,t)\quad&=\quad\cases{\phi_A^{(i)}(\vec x,t)&$t>0$\cr
                                                  \phi_A^{(i+1)}(\vec x,t)&$t<0$}  \cr
\tilde\phi_B^{(i)}(\vec x,t)\quad&=\quad\cases{\phi_B^{(i)}(\vec x,t)&$~~~t>0$\cr
                                                  \phi_B^{(i)}(\vec x,t)&~~~$t<0$}  \cr
}}
The fields $\tilde\phi^{(j)}$ are single valued at $t=0$ by construction.

However there is a new ``gluing'' condition due to the fact that $\phi^{(i)}$ was single valued at the
boundary $\p$ between $A$ and $B$, i.e.
\eqn\newgluing{
\phi^{(i)}_A(\vec x,t)\Big|_{\vec x\in\p}=\phi^{(i)}_B(\vec x,t)\Big|_{\vec x\in\p}}
With the rearrangement \newphi, \newgluing\ becomes
\eqn\finalgluing{
\tilde\phi_A^{(i)}(\vec x,t)\Big|_{\vec x\in\p}=\tilde\phi_B^{(i+1)}(\vec x,t)\Big|_{\vec x\in\p}
\qquad\hbox{for $t<0$}}
So the new formulation is:\hfill\break
a set of $n$ replicas $\tilde\phi^{(i)}$ with a total action on single valued fields
\eqn\newaction{
S=\sum_i S[\tilde\phi^{(i)}]}
and the gluing of different fields \finalgluing.
The boundary conditions at $t=\pm\infty$ are left free for each $\tilde\phi^{(j)}$.

Following \Cardy\ we will now `uniformize' the gluing manifold.
The CFT is defined on ${\R}^{3,1}$ with the flat euclidean metric ($x^4=t$)
\eqn\metricRfour{
ds^2=(dx^1)^2+(dx^2)^2+(dx^3)^2+(dx^4)^2}
In the vicinity of the boundary $\p$ we choose Gaussian normal
coordinates $(x^i)|_{i=1,2,3}\to(r,y^a)$, $a=1,2$, such that the boundary is
located at $r=0$ and the metric is of the form
\eqn\metricnew{ ds^2=dt^2+dr^2+g_{ab}(r,x)dy^a dy^b}
The induced metric on the $r=0$ hypersurface (at constant $t$) is $h_{ab}=g_{ab}|_{r=0}$.
The components of the second fundamental form
$K^r_{ab}=-{1\over2}\p_r g_{ab}|_{r=0}$ are generically non-zero.
We remark that the choice of Gaussian coordinates leading to the metric
\metricnew\ is also possible if we replace ${\R}^{3,1}$ by ${\cal M}\times {\R}$, where
${\cal M}$ is any three-manifold. The induced metric and second fundamental
form are still expressed in terms of $g_{ab}$ as above.

The gluing manifold is now
\eqn\glingmfd{
r=0,\,t<0}
A further (singular) change of variables
\eqn\uniform{
w\equiv it+r ={z^{n}}}
brings all the gluing regions to the $z$ plane times $\p$. This is exemplified for $n=3$
in the figure which shows the $z$-plane.
\bigskip
$~~~~~~~~~~~~~~~~~~~~~~~~~~~~~~~~~~~~~~~~~~~~~~~~~~~~~~
~~~~~~~~~~\tilde \phi^{(1)}$
\vskip.2cm
$~~~~~~~~~~~~~~~~~~~~~~~~~~~~~~~~~~~~~~~~\tilde\phi^{(2)}$
\vskip.8cm
$~~~~~~~~~~~~~~~~~~~~~~~~~~~~~~~~~~~~~~~~~~~~~~~~~~~~~~
~~~~~~~~~~\tilde \phi^{(3)}$
\vskip-3.2cm
\figinsert\figone{}{1.2in}{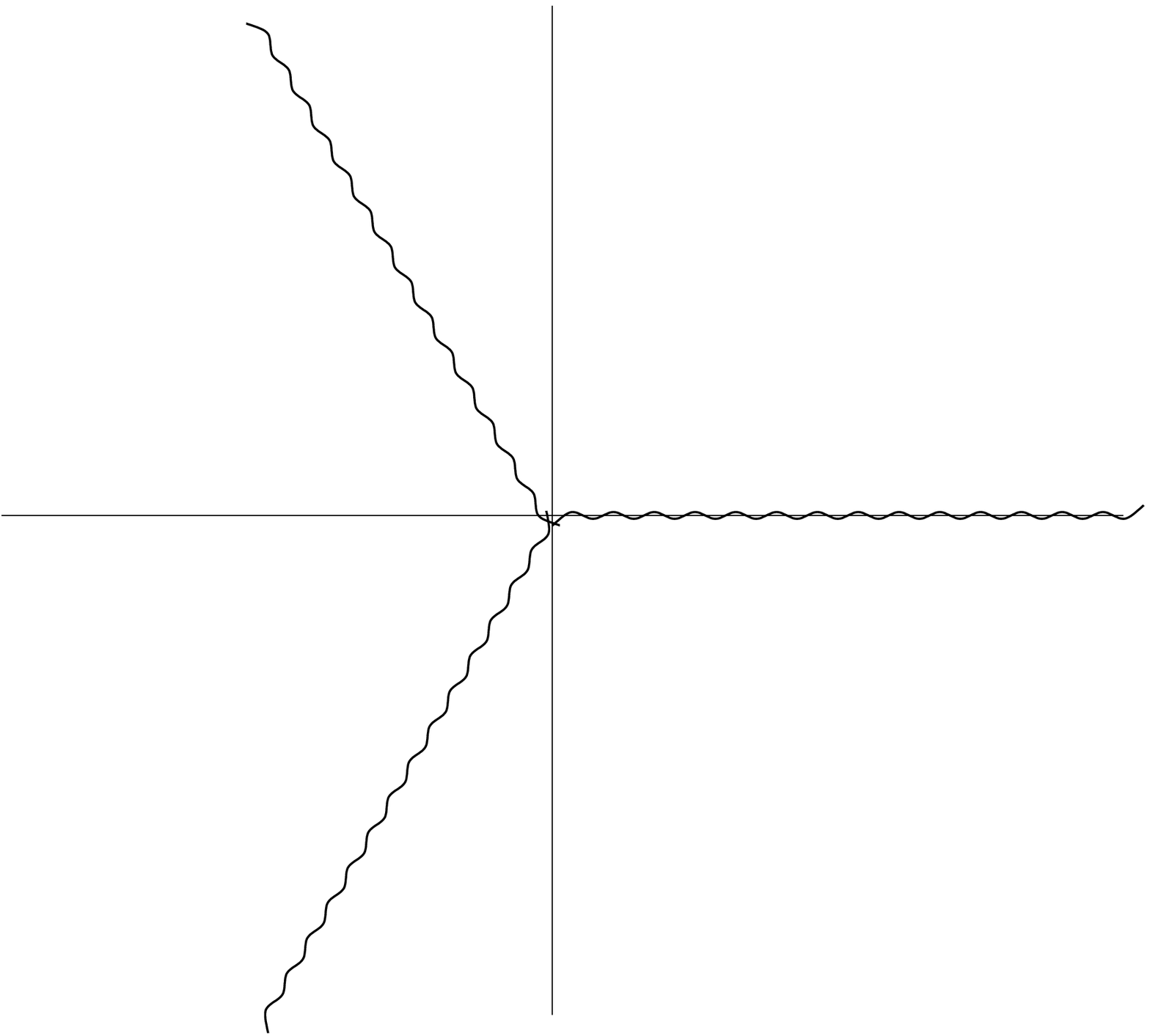}
\vskip-1.2cm
\noindent

The metric becomes:
\eqn\metricchange{ ds^2=n^2(z\bar z)^{(n-1)}dz d\bar z+g_{ab}(r,x)dy^a dy^b}
For $n\neq 1$ this metric is singular on the hypersurface $z=0$ which is the
boundary between the regions $A$ and $B$ at time $t=0$. This
singularity is a consequence of the singularity of the coordinate
transformation \uniform.

Since we are going to study what is happening when the metric
undergoes a space-dependent Weyl transformation we allow
\metricchange\ to be multiplied by such a factor such that finally
the most general class of metrics we are studying is
\eqn\genmet{ ds^2= 2 g_{z\zb}(r,x)(z\zb)^{(n-1)}dzd\zb
+g_{ab}(r,x)dy^a dy^b}
where $g_{z\zb}$ is proportional to the respective metric component
before the singular diffeomorphism.
\newsec{Holographic anomalies}

In this section we present a general scheme to compute holographic
conformal anomalies. It is very much like the computation of the
$SU(4)$ ${\cal R}$-current anomaly presented in \WittenI. The
anomaly is the boundary term generated by a suitably chosen local
symmetry transformation. In the case of the ${\cal R}$-current this
is a $SU(4)$ gauge transformation and the boundary term, which is
the ${\cal R}$-current anomaly of the CFT on the boundary
(${\cal N}=4$ SYM) is due to the $SU(4)$ Chern-Simons term present in the 5d
gauged supergravity that arises when one compactifies type IIB
string theory on $S^5$. For the conformal anomalies the appropriate
transformations are the so-called PBH-transformations, a subgroup
of five dimensional diffeomorphisms introduced in \ISTY\ and
reviewed in the following. This treatment of the conformal
anomalies does not require the solution of the equations of motion
and it does not depend on the introduction of a cutoff. This gives
us confidence on the generality of their structure when we compare
it with the results in the field theory.

In the first part of this section we deal with those anomalies
origination from the bulk gravitational action. In the second part
we extend the discussion to the trace anomalies originating from the
DNG piece of the action (``Graham-Witten anomalies'').

\subsec{Anomalies from the bulk}

We start with trace anomalies in the bulk. Besides giving a general
illustration of the new way to calculate trace anomalies the
explicit results will be used in the following for an alternative
holographic representation of the anomalous pieces in the EE.

Consider a generic gravitational bulk action
\eqn\bulkaction{
S=\int_M\sqrt{G}f(R)\,d^{d+1}X}
where $f$ is an arbitrary {\it scalar} function  of the curvature
and its derivatives. We require that \bulkaction\ admits $AdS_{d+1}$
as a solution to the equations of motion: this imposes a mild
inequality on the coefficients in  $f(R)$.

We choose coordinates $X^\mu=(x^i,\rho)$ such
that $\rho=0$ is the boundary of $AdS_{d+1}$ where the dual CFT
lives. It is coupled to a metric (source for its energy momentum
tensor) $g^{\sss(0)}_{ij}(x)$. For the bulk metric we choose the
Fefferman-Graham (FG) gauge \FG\HS
\eqn\FGmetric{
ds^2=G_{\mu\nu}\,dX^\mu dX^\nu
=\left({d\rho\over 2\rho}\right)^2+{1\over\rho}g_{ij}(x,\rho)dx^i dx^j}
with $g_{ij}(x,0)=g^{\sss(0)}_{ij}(x)$. For $g^{\sss(0)}_{ij}=\eta_{ij}$ \FGmetric\ is the
metric of $AdS_{d+1}$, whose curvature radius we have set to one.

PBH (Penrose-Brown-Henneaux) transformations are those
diffeomorphisms $\xi^\mu$ which preserve the FG-gauge \ISTY, i.e.
for which ${\cal L}_\xi G_{\rho\rho}={\cal L}_\xi G_{\rho i}=0$. The
solution is parametrized by an arbitrary function $\s(x)$:\foot{The
choice of lower limit in the $\rho'$ integral means that we do not
consider diffeomorphisms of the boundary. They are of no interest
here.}
\eqn\PBH{
\xi^\rho=-2\rho\s(x)\,,\qquad
\xi^i=a^i(x,\rho)={1\over2}\int_0^\rho\,d\rho' g^{ij}(x,\rho')\p_j\s(x)}
In particular, $\delta_\s g^{\sss(0)}_{ij}=2\s g^{\sss(0)}_{ij}$,
i.e. $\s(x)$ is the parameter of Weyl rescalings of the boundary metric.

The group property for PBH transformations can be shown to be
\eqn\groupproperty{
\xi_1^\nu \p_\nu \xi_2^\mu-\xi_2^\nu\p_\nu\xi_1^\mu+\delta_2\xi_1^\mu-\delta_1\xi_2^\mu=0}
The last two terms are due to the dependence of the transformation parameters
on $g_{ij}(x,\rho)$.

The essential property of the PBH transformations is that on the
boundary they coincide with the action of the Weyl group. Therefore
in holography the Weyl group becomes embedded in the $d+1$
dimensional diffeomorphisms  and the study of Weyl anomalies is
reduced  to an analysis of how diffeomorphisms act.

Under a bulk diffeomorphism the action \bulkaction\ is invariant up
to a boundary term
%
%
%
\eqn\transL{
\delta_\xi (\sqrt{G}f)=\p_\mu(\sqrt{G}\xi^\mu f)}
\eqn\boundaryterm{
\delta_\xi S=\int_{\partial M}d^d x\sqrt{G}f(R)\,\xi^\rho|_{\rho=0}
=-2\int_{\partial M}d^d x\sqrt{G}f(R)\,\rho\,\sigma|_{\rho=0}}
where in the second line we have restricted
the diffeomorphism to a
PBH transformation. The finite piece of this boundary term is the
holographic Weyl anomaly.

A comment is in order here: we consider passive diffeomorphism transformations
which act on the fields rather than the coordinates. The reason for doing is that we
want to keep the boundary fixed.

Following \FG (see also \AR)
we expand the metric as
\eqn\FGexpansion{
g_{ij}(x,\rho)=\sum_{n=0}^\infty \gn_{ij}(x)\rho^n+\dots}
The $\dots$ denote logarithmic terms ($\sim\log\rho$) which are present for even $d$.
They do not play a role in our analysis.
The integrand in \bulkaction\ has likewise an expansion of the form \ISTY
\eqn\PBHaction{
\sqrt{G}f(R)=\sqrt{\textstyle\gz}\rho^{-{d\over2}-1}b(x,\rho)
=\sqrt{\textstyle\gz}\rho^{-{d\over2}-1}\sum_{n=0}^\infty b_n(x)\rho^n}
As was shown in \ISTY, $b$ and thus each $b_n$,
satisfies the Wess-Zumino consistency condition
\eqn\WZbn{
\int d^d x\sqrt{g^{\sss{(0)}}}(\s_1\delta_{\s_2}-\s_2\delta_{\s_1})b=0}
A simple way to see this is as follows (cf. the Appendix).
For ${\cal O}=\sqrt{G}f(R)$ one
derives $\delta_{\s_1}{\cal O}=\p_\mu(\xi_1^\mu{\cal O})$
and $[\delta_{\s_2},\delta_{\s_1}]\,{\cal O}=0$ by virtue of the group property
\groupproperty.

On-shell $b_n$ is a local, covariant expression constructed from
$g^{\sss(0)}_{ij}$. For $d=2n$ it is the coefficient of the boundary
term at ${\cal O}(\rho^{-1})$ and represents the Weyl anomaly of the
dual $2n$-dimensional CFT. The bulk gravitational action thus plays
the same role for the Weyl anomaly of the CFT as does the CS term
for the ${\cal R}$-current anomaly.

For general $d=2n$, the on-shell $b_n$ depends also on some of the
derivatives of $\g_{ij}$ at $\rho=0$ and not only on the boundary
value $g^{\sss(0)}_{ij}$. These higher derivatives need some
information contained in the equation of motion. However, for
$d=4$, $b_2$ can be computed without the need to solve the equations
of motion. As we will show momentarily, in $d=4$, besides $\gz$ only the
coefficient $\go$ of the FG expansion of the bulk metric appears.
This second coefficient is universal because it is uniquely
determined by its behavior under PBH transformations and locality
\ISTY:
\eqn\gone{
\go_{ij}={1\over(d-2)}\Big(R_{ij}-{1\over2(d-1)}g_{ij}R\Big)}
where $R$ is the curvature of $g^{\sss(0)}$ and $g_{ij}\equiv
g^{\sss(0)}_{ij}$. The universality of $\g^{\sss(1)}$ will be spoiled if
we take back reaction into account, as we will do in section 5.

On dimensional grounds $b_n$ can at most be linear in $\g^{\sss(n)}$, as
both carry length-dimension $-2n$ ($\rho\sim{\rm length}^2$). By
assumption, $f(R)$ is such that Anti-de-Sitter space is a solution
of the equations of motion. Expand the action around this solution.
In this expansion the term linear in the fluctuations around the
$AdS$-metric can only be a total derivative (or vanish altogether).
Consider the terms $\nabla^\mu\nabla^\nu \delta G_{\mu\nu}$ and
$\boxx \tr\delta G$. For fluctuations $\delta
G_{ij}=\rho^{n-1}g^{\sss(n)}_{ij}$ the possibly dangerous terms, i.e.
those which might contribute to $b_n$, are of the type $\rho^n
\tr\,g^{\sss(n)}$. It is straightforward to show that their coefficient
is zero for $d=2n$. Higher derivative terms in the variation of the
action will involve coefficients $g^{\sss(m)}$ for $m<n$. We stress that
the above argument showing that in $d=2n$ $g^{\sss(n)}$ does not appear in
the $b_n$ term in the expansion of the action does not prevent the
participation of $g^{\sss(n)}$  in the equation of motion in the usual way
\HS\ of calculating the anomalies.\foot{Alternative arguments for 
obtaining anomalies without solving the equations of motion were given in 
\ST\PS.}

To summarize, to find the Weyl anomaly of the $d=2n$-dimensional
dual CFT all we have to do is to extract the coefficient of $1/\rho$
is the expansion of the gravitational action. In $d=4$ this only
involves $\gz$ and $\go$ and is thus completely fixed. On general
grounds this can always be written as a linear combination
$a\, E_4-c\, C^2 +e\, \boxx R$ where $C^2$ is the square of the Weyl tensor, $E_4$
the Euler density (i.e. $\int_M\sqrt{g} E_4\propto \chi(M)$). In the
Appendix we will rederive the general expression for $a$, already
found, by different means, in \ISTY.

\subsec{Graham-Witten anomalies}

In this subsection we will study the trace anomalies  for
submanifolds (``Graham-Witten'' anomalies) which are of direct
relevance  for the proposed holographic dual of EE. We will follow
the method used  in the previous subsection for bulk anomalies which
does not depend on the equations of motion and does not need a
cutoff. This will enable us to discuss the general structure of the
Graham-Witten anomalies needed for EE and  the anomalies produced by
more general submanifold actions having the same symmetries  as DNG.
For  the DNG action our method reproduces the result of \GW.

We start with a classification of the possible Graham-Witten anomalies for
the case when the submanifold  has dimension $2$ embedded in
a manifold of dimension $d$.

Candidates for the Graham-Witten anomaly are solutions to the
Wess-Zumino consistency condition satisfying the following
conditions: they should be local expressions constructed from the
second fundamental form and from  curvatures, linear in the Weyl
parameter $\sigma$; they should have two derivatives (appropriate
for the two dimensional case considered here); they should be
cohomologically non-trivial. Among those we distinguish between type
A which satisfy the WZ condition non-trivially and type B which
satisfy them trivially having expressions which are Weyl invariant.

To find the candidates for the anomaly, we will need, besides well-known expressions
for the Weyl-transformation of the curvature tensors, the transformation of the
second fundamental form and of its trace (cf. below):
\eqn\deltaK{ \delta_\s K^i_{ab}=-h_{ab}P^{ij}\p_{j}\s\,,\qquad
\delta_\s K^i\equiv \delta_\s(h^{ab}K^i_{ab})=-2\s K^i-k\, P^{ij}\p_j\s}
where $P^{ij}=g^{ij}-h^{ij}=g^{ij}-h^{ab}\p_a X^i\p_b X^j$
projects to the normal space of the hypersurface. The derivatives 
$\p_a$ are with respect to the local coordinates on the submanifold
and $X^i$ are the embedding functions. 

It is then straightforward to show that the following list exhausts all
possible Weyl invariant expressions:
\eqn\listWi{\sqrt{h}h^{ac}h^{bd}C_{abcd},\quad
\quad \sqrt{h}(\tr(K^i K^j)-{\textstyle 1\over2}K^i K^j)g_{ij},\quad
\sqrt{h}(K^i K^j g_{ij}-4h^{ab}\go_{ab}+2 R^{(2)})}
where $R^{(2)}$ is the curvature scalar of the induced metric, $C_{abcd}$ the pull-back
of the bulk Weyl tensor and
$\displaystyle{\go_{ab}=\p_a\!\Xz^i\,\p_b\!\Xz^j\go_{ij}}$ 
is the pull-back of \gone.
However, with the help of the Gauss-Codazzi equation one shows that
\eqn\useGC{
h^{ac}h^{bd}C_{abcd}=R^{(2)}-2h^{ab}\go_{ab}+\textstyle{1\over2}K^i K^j g_{ij}
-(\tr(K^i K^j)-{1\over2}K^i K^j)g_{ij}}
i.e. the above Weyl invariant expressions are not all independent.
We will choose the first two as a basis.

Candidates for the type A anomaly are
\eqn\listA{ \sqrt{h}R^{(2)}\s,\quad\sqrt{h}K^i\p_i\s,\quad\sqrt{h}\boxx \s}
where these expressions are restricted to the submanifold.
The first is the well-known trace anomaly in $d=2$. The second
one, on the other hand, is trivial as it can be written as the Weyl
variation of a local term:
\eqn\trivial{
\delta_\s(K^i K^j g_{ij})=-4 K^i\p_i\s}
where one uses \deltaK\ and $K_i h^{ij}=0$. The third one is again
trivial being the variation of $R$ the scalar bulk curvature
restricted to the submanifold.

We thus arrive at
the following basis of GW anomalies when the submanifold is two
dimensional:
\eqn\classificationAB{
\eqalign{
\hbox{type A:}&\quad \sqrt{h}R^{(2)}\s \cr
\hbox{type B:}&\quad \sqrt{h}h^{ac}h^{bd}C_{abcd}\,\s,\quad
\sqrt{h}g_{ij}(\tr(K^i K^j)-\textstyle{1\over2}K^i K^j)\,\s}}
In terms of this basis the anomaly found by Graham and Witten, who considered
the case where the hypersurface degrees of freedom in the CFT have their
holographic description in terms of the DNG action, is
\eqn\GWAB{
{\cal A}_{GW}={1\over4}\int_{\p\Sigma}\!\! d^2 x\,\sqrt{h}
\Bigl(h^{ac}h^{bd}C_{abcd}-g_{ij}(\tr(K^i K^j)-\textstyle{1\over2}K^i K^j)-R^{(2)}\Bigr)\s}

We proceed now to an analysis of the Graham-Witten anomalies in a
holographic setup. We will leave the dimensions of space-time $d$
and of the submanifold $k$ general  and at the end of the discussion
we will go back to the specific $k=2 $ case.

In the holographic realization we have to consider  a
$(k+1)$-dimensional submanifold $\Sigma$ embedded into the
$(d+1)$-dimensional bulk $M$ such that it ends on a $k$-dimensional
submanifold $\p\Sigma$ on the $d$-dimensional boundary. Denote, as
before, the bulk coordinates by $X^\mu=(x^i,\rho)$ and the
world-volume coordinates by $\tau^\a=(y^a,\tau)$ with $i=1,\dots,d$
and $a=1,\dots,k$. The embedding is $X^\mu:\, \Sigma\mapsto M$, i.e.
$X^\mu=X^\mu(\tau^\a)$.

We assume that the action contains in addition to the usual bulk
component \bulkaction\ another component defined on the $k+1$
submanifold. The additional piece is invariant both under usual
bulk diffeomorphisms and under reparametrizations of the world
volume.

We want first to generalize the PBH transformations \PBH\ to this
new situation where we have two linked gauge invariances.

We first fix the gauge. For the bulk we go to FG gauge \FGmetric\ as
before. The reparametrizations  of $\Sigma$ are fixed by imposing
\eqn\gfixSigma{
\tau=\rho\qquad{\rm and}\qquad h_{a\tau}=0}
Under a reparametrization of $\Sigma$, parametrized by
$\tilde\xi^\a$, $X^\mu$ transforms as a scalar, i.e.
$\delta_{\tilde\xi}X^\mu=\tilde\xi^\a\p_\a X^\mu$. In particular
$\delta_{\tilde\xi}\rho=\tilde\xi^\a\p_\a\rho=\tilde\xi^\tau=0$
after fixing the $\tau=\rho$ gauge. Also, if we require that
$\delta_{\tilde\xi}h_{a\tau}=0$, we find that $\tilde\xi^a$ must be
independent of $\tau$. This means that all world-volume
reparametrizations of $\Sigma$ are fixed except the ones acting on
$\p\Sigma$.

We perform now a target space PBH transformations
$\delta\rho=-2\rho\s,\,\delta x^i=a^i$ (cf. \PBH). To stay in the
$\tau=\rho$ gauge we must make a compensating world-volume
diffeomorphism
\eqn\compwd{
\tilde\xi^\tau=-2\tau\sigma}
The resulting change of the induced metric must be compensated in
order to keep $h_{a\tau}=0$:
\eqn\deltah{
\delta h_{a\tau}=\p_a\tilde\xi^\tau h_{\tau\tau}+\p_\tau\tilde\xi^b h_{ab}=0}
With $\tilde\xi^\tau=-2\tau\sigma$ this can be integrated to
\eqn\tildexi{
\tilde\xi^a=2\int_0^\tau\!d\tau'~\tau' h_{\tau\tau}h^{ab}\p_b\s}
where all functions in the integrand depend on $\tau'$ (through $X^i(y^a,\tau)$).
Here
\eqn\htautau{
h_{\tau\tau}=\p_\tau X^\mu\p_\tau X^\nu G_{\mu\nu}={1\over4\tau^2}
+{1\over\tau}\p_\tau X^i\p_\tau X^j g_{ij}(X,\tau)}
Expand $g_{ij}$ in powers of $\rho$ (cf. \FGexpansion) and $X^i$ in powers of
$\tau$ (with $\tau=\rho$)
\eqn\xiexp{
X^i(\tau,y^a)=\Xz^i(y^a)+\tau \Xo^i(y^a)+\tau^2\Xt^i(y^a)+\dots}
With the definition
\eqn\defhzero{
h_{ab}={1\over\rho}\p_a X^i\p_b X^j g_{ij}(X)
={1\over\rho}\p_a\!\!\Xz^i\,\p_b\!\!\Xz^j\gz_{ij}(\Xz)+{\cal O}(1)
\equiv {1\over\rho}\hz_{ab}(X)+{\cal O}(1)}
we obtain from \tildexi\
\eqn\tildexiexp{
\tilde\xi^a={1\over2}\tau \hz^{ab}\p_b\s+{\cal O}(\tau^2)}
We can now determine how $X^i$ changes under PBH. It transforms as
\eqn\deltaPBHXi{
\delta X^i=\tilde\xi^\a\p_\a X^i-a^i}
%
with $a^i$ from \PBH.
This implies
\eqn\deltaXzXo{
\eqalign{
\delta \Xz^i&=0\cr
\delta \Xo^i&=-2 \s \Xo^i+{1\over 2}\hz^{ab}\p_a\!\!\Xz^i\,\p_b\s-{1\over2}\gz^{ij}\p_i\s}}
which is solved by
\eqn\solXo{
\Xo^i={1\over2k}K^i}
where
\eqn\defKi{
K^i=\hz^{ab}K^i_{ab}=\hz^{ab}\Bigl(\p_a\p_b\!\!\Xz^i-\Gz_{ab}^c\p_c\!\!\Xz^i
+\Gz_{jk}^i\p_a\!\!\Xz^j\p_b\!\!\Xz^k\Bigr)}
is the trace of the second fundamental form, i.e. the extrinsic curvature,
of the embedded submanifold $\p\Sigma$.

We remark that the universality of ${\displaystyle \Xo^i}$ is
analogous to the universality of $g^{\sss(1)}$, c.f. \gone. The
higher $\Xn$, just like the higher $g^{\sss(n)}$, are not universal, the
reason being that their behavior under PBH transformations admits
homogeneously transforming terms \foot{For $g^{\sss(2)}_{ij}$ this
is e.g. $g^{\sss(0)}_{ij}C^2$ and for $X^{\sss(1)}$ any one of the terms
in \listWi\ (without the $\sqrt{h}$ factor), multiplied by
$X^{\sss(0)}$.}.

We succeeded therefore to put also this more general
situation, with the action having two components, into a framework
similar to the one we had for the bulk action alone.
The action of the Weyl transformations on the boundary is embedded
into bulk diffeomorphisms and world volume reparametrizations \compwd,
\tildexi. Moreover, besides the $g^{\sss(1)}$ component of the bulk
metric also the $X^{\sss(1)}$ component of the embedding have a
universal form determined by the PBH transformations.

Using these results we are now  prepared to analyze the
Graham-Witten anomalies, i.e. the
transformation properties of the additional piece of the action
when the metric $g^{\sss(0)}$ is Weyl transformed.

Following \GW\ we consider the case where the dynamics of the
submanifold is governed by the DNG action:
\eqn\DNG{
S=\int_\Sigma\sqrt{h}}
The generalization to
arbitrary world-volume actions is straightforward. A particular case will be
considered at the end of this section.
The DNG action of $\Sigma$ is invariant under passive
world-volume diffeomorphisms up to a
boundary term. The finite part of this boundary term (at $\tau=0$)
is the Graham-Witten anomaly
\eqn\GWanomaly{
{\cal A}=\int_{\p\Sigma}\!\!\sqrt{\det h}\,\tilde\xi^\tau|_{\rm finite}}
Given that the $\tau$-expansion of $\Xo$ is universal only up to the first non-trivial
order, we will be able to compute the anomaly, without further input from the equations
of motion, only for $k=2$. This is
also the relevant dimension for the discussion of the EE
in a four dimensional CFT.

We now evaluate \GWanomaly. We need
\eqn\xitildetau{
\tilde\xi^\tau=-2\tau\s(X)=-2\tau\s(\Xz)-2\tau^2\p_i\s(\Xz)\Xo^i+{\cal O}(\tau^3)}
and $\det(h)=h_{\tau\tau}\det(h_{ab})$ with
\eqn\deth{
\eqalign{
h_{\tau\tau}&=
{1\over4\tau^2}+{1\over\tau}\Xo^i\Xo^j\gz_{ij}+\dots={1\over4\tau^2}(1+4\tau \Xo^i\Xo^j\gz_{ij})+\dots\cr
\det h_{ab}&={1\over\tau^k}\det(\hz_{ab})(1+\tau\hz^{ab}\ho_{ab})+\dots}}
where
\eqn\hzeroab{
\eqalign{
\ho_{ab}&=
\p_a\!\!\Xo^i\,\p_b\!\!\Xz^j\gz_{ij}+\p_a\!\!\Xz^i\,\p_b\!\!\Xo^j\gz_{ij}
+\p_a\!\!\Xz^i\,\p_b\!\!\Xz^j\go_{ij}+\p_a\!\!\Xz^i\p_b\!\!\Xz^j\p_k\!\!\gz_{ij}\Xo^k\cr
&=\go_{ab}-{1\over k}K^i K^j_{ab}\gz_{ij}}}
With the help of these expressions we finally find, for $k=2$,
\eqn\GWktwo{
{\cal A}_{GW}={1\over8}\int_{\p\Sigma}\!\! d^2 y\,\sqrt{\det h}
\Bigl(\big(g_{ij}K^i K^j-4 h^{ab}\go_{ab}\big)\s-2K^i\p_i\s\Bigr)}
where $h$ and $g$ now denote the boundary metrics. Eq.\GWktwo\ is in
agreement with \GW. As remarked above, the last term is cohomologically trivial.
The rest can be written in terms of the basis \classificationAB.
The result was already given in \GWAB.

In analogy to allowing general bulk actions, as we did in Section 3.1,
the dynamics of the hypersurface might be given by generalizations of the DNG action
\eqn\DNGGEN{
S=\int_\Sigma\sqrt{h}f(R^{\Sigma},K,X,\dots)}
where $f$ is a scalar function.
In this case the GW anomaly will also change:
\eqn\GWgen{
{\cal A}=\int_{\p\Sigma}\sqrt{\det h}f\tilde\xi^\tau|_{\rm finite}}
The fact that the anomaly satisfies the WZ condition is again a consequence of the
group property of the PBH transformations.

As a particular example we consider the action
\eqn\DNGgen{
S=\int_\Sigma\sqrt{h}f(R^{(\Sigma)})}
where $R^{(\Sigma)}$ is the Ricci scalar computed with $h_{\a\b}$
with the expansion
\eqn\tauexpR{
R^{(\Sigma)}=6+\left(R^{(2)}-2 \hz^{ab}\go_{ab}+{\textstyle{1\over2}}\gz_{ij}K^i K^j\right)\tau+\dots}
For instance, if we choose $f(R^{(\Sigma)})=1-{1\over2}R^{(\Sigma)}$
the GW-anomaly is purely type A. Alternatively we can choose  an
action for which the $R^{(2)}$ anomaly vanishes which could be
relevant for the EE as we discuss in Section 5.

\newsec{Anomalies in the singular metric background}

We will study the CFT in the singular metric \genmet\ obtained in
Section 2 to which the CFT should be coupled in order to calculate
the EE. The metric is the result of a singular diffeomorphism
\uniform\ applied to the original, smooth metric.

We will assume that in the generating functional the singular
metric can be used for calculating the EE at
least for the terms generating the trace
anomalies without the need for additional regularization.

Since through the ``replica trick'' the EE
requires only the derivative with respect to the replica number
$n$ at $n=1$ we will expand everything  in $\epsilon=n-1$
and keep only the first order term in $\epsilon$.

The thus expanded, singular diffeomorphism is given by:
\eqn\diffeo{w=z+\epsilon z \log(z)\qquad
\bar w=\zb+ \epsilon \zb \log(\zb)}
As a consequence of the diffeomorphism being singular there could be two
dimensional $\delta$- function contributions in certain curvature
components. The contribution which can give $\delta$-function
appears in $g_{z\zb}$:
\eqn\sing{\epsilon g_{z\zb} \log (z\zb)}
Using
\eqn\delt{\partial_{z}\partial_{\zb}\log(z\zb)=4\pi\delta^{(2)}(z,\zb)}
the singular contribution is:
\eqn\curv{\bar R_{z\zb z\zb}=-4\pi \epsilon g_{z\zb}\delta^{(2)}(z,\zb)}
where we denote with $\bar R{....}$ the contributions to the
transformed curvature components  which have a $\delta$-function. In
addition the transformed components will have contributions denoted
by $\tilde R{...}$ through the action of \diffeo\ treated as a
regular diffeomorphism on the original components $R{...}$.

We list the components of the Ricci tensor and the scalar
curvature which are singular as a consequence of \curv:
\eqn\ricci{\bar R_{z\zb}=4 \pi \epsilon \delta^{(2)}(z,\zb)
\qquad \bar R=8\pi \epsilon
g^{z\zb}\delta^{(2)}(z,\zb)}
It is tempting to separate the ``regular'' and ``singular'' pieces
of the curvatures representing the EE to give respectively a four
dimensional theory in a smooth metric  and an effectively two
dimensional contribution obtained from the singular piece after
integrating the $\delta$-function. Such a separation would justify
the holographic representation proposed in \RT\ where in addition to
the five dimensional bulk theory there is the DNG action
representing the two dimensional, singular contribution.

However, as discussed in Section 3 the aforementioned holographic
setup leads to a well defined, specific analyticity structure. In
order   that the holographic mapping makes sense the same
analyticity structure should exist in the original theory. In
particular the holographic representation leads to an effective
action depending on the boundary variables (metric and embedding
functions of the submanifold) which is Weyl invariant up to local
anomalies. This means that:

a) under a Weyl transformation the effective action is invariant
except for local terms: four dimensional (``the bulk trace
anomalies'') and two dimensional (``the Graham-Witten anomalies'')
and

b) the anomalies fulfill the Wess-Zumino condition, i.e. a
further Weyl variation antisymmetrized with the first one should
vanish.

We will study the above conditions for the universal pieces
of the effective action of the EE field theory responsible for the
trace anomalies and show that they fail  for rather generic
geometries. We believe that similar failures probably occur also in
other terms of the effective action which however are specific to
the various CFT.

The calculation  is rather straightforward. We introduce the
singular metric to first order in $\epsilon$  into the effective
action and the Weyl anomalies obtained from it by making a Weyl
variation. The expressions make sense to first order in $\epsilon$
in the four dimensional sense. We will try, however, to ``split'' the
expressions into a four dimensional piece corresponding to the
regular part of the curvatures and a two dimensional piece
contributed by the components of the curvatures which have an
explicit $\delta$-function and we will check if the two pieces have
the properties of the holographic representation.

If the splitting is done at the level of the effective action this
requires that after a Weyl variation the two pieces produce local
Weyl anomalies, bulk (four dimensional) and Graham-Witten (two
dimensional), respectively.

If the splitting is done for the Weyl anomalies in the background of
the singular metric (again to first order in $\epsilon$) the two
pieces should obey the Wess-Zumino conditions in four and two
dimensions, respectively.

Both ``splittings'' fail   for the part of the effective action
producing the type A (Euler) anomaly if the embedding geometry has a
nonvanishing second fundamental form. An analysis of the trace
anomalies for the EE when the second fundamental form vanishes was
performed in \RT\ using the geometric setup discussed in \FS.

Since the analysis of the nonlocal anomalous pieces of the action is
somehow cumbersome we will use the Wess-Zumino actions \WZ\ in
which only manipulation of local terms is needed the results being
completely equivalent. The Wess-Zumino actions replace the
nonlocality with the introduction of another scalar field
$\phi(x^i)$, the parameter field of the Weyl group. The
dependence of the action  on the metric $\g_{ij}$ and on $\phi$
is local. The Weyl transformation of the fields is:
\eqn\trans{g'_{ij}(x^{k})=\exp(2\sigma(x^{\k})) g_{ij}(x^{k}) \qquad
\phi'(x^{k})=\phi(x^k) + \sigma(x^k)}
where $\sigma(x^{k})$ is the parameter of the Weyl
transformation. We require that the effective action
$W(g_{ij},\phi)$ under the transformation \trans\ produces the
anomalies, i.e.
\eqn\wzact{\delta_{\sigma} W=\int d^d x\, \sigma \A}
where the anomaly $\A$ depends only on $g_{ij}$.
The effective actions $W$ are obtained by a general procedure \WZ\ and
we reproduce here their form in $d=2$ and $d=4$ we will need:
\eqn\wzt{W_2=a_2 \int d^{2}x \sqrt{\det(g)}[\phi R+g^{ab}\partial_{a}\phi\partial_{b}\phi]}
where $a_2$ is the unique two dimensional trace anomaly coefficient, and
\Cappelli
\eqn\wzf{W_4=\int d^{4}x \sqrt{\det(g)}\Big( a_4 [\phi E_{(4)}
-4 G_{ij} \partial^{i}\phi \partial^{j}\phi
+2(\partial_{i}\phi \partial^{i}\phi)^{2}
-4\boxx\phi\partial_{i}\phi\partial^{i}\phi] +c_4 \phi
C^{2}_{(4)}\Big)}
where $E_{(4)}$ is the four dimensional Euler density,
$C^{2}_{(4)} $ is the square of the four dimensional Weyl
tensor, $G_{ij}$ is the Einstein tensor
\eqn\eins{G_{ij}=R_{ij}-\textstyle{1\over2}g_{ij}R}
and $a_4$ and $c_4$ are the coefficients of the two four dimensional trace anomalies.

To the expressions \wzt,\wzf\ we can add arbitrary terms made
of $\phi$ and $\g$ which are Weyl invariant, in particular the
quadratic action of a conformally coupled scalar $\Phi$,
the exponential of $\phi$.

If we express $\phi$ through the equation of motions in terms of $g$
and use it in \wzt\ or \wzf\ the equation \wzact\ is still fulfilled,
i.e. we obtain the nonlocal action generating the anomaly. We
illustrate this procedure in $d=2$: The $\phi$ equation of motion
gives:
\eqn\eqm{\phi={1 \over 2\boxx} R}
and using \eqm\ in \wzt\ we obtain:
\eqn\poly{W_2 = {a_2 \over 4}\int d^2 x \sqrt{\det (g)} R {1\over \boxx} R}
i.e. the Polyakov action.

A similar procedure can be done in $d=4$ solving  the
equation of motion following from \wzf\ as an expansion in
powers of $\phi$.
We will not need the explicit form of the
nonlocal action obtained this way being more convenient to
work with the local action \wzf\ containing $\phi$.

We will try to implement the ``splitting'' mentioned above by using
the singular and regular components of the Riemann tensor in \wzf.
The terms which could have $\delta$-functions are the ones
containing $E_{(4)}$,  $G_{ij}$ and $C^2_{(4)}$. There are no
singular contributions from the $\phi$ field since the derivatives
acting on it are not of a high enough order to produce a
$\delta$-function.

We start with the calculation of the piece of \wzf\ responsible for
type A (Euler) anomaly. This requires evaluating the first two
expressions  to order $\epsilon$.

Since in $d=4$ $E_{(4)}$ is given by:
\eqn\euler{E_{(4)}=\textstyle{1\over4} \epsilon^{i_1 i_2 j_1 j_2}
\epsilon^{i_3 i_4 j_3 j_4} R_{i_1 i_2 i_3 i_4}
R_{j_1 j_2 j_3 j_4}}
the singular curvature component \curv\ will single out the
regular component  $R_{abcd}$ where the indices $a,b,c,d$ take
the values $1,2$. The singular components of the Einstein tensor
using \ricci\ will be in the $1,2$ directions only, i.e.
\eqn\einsing{\bar G_{ab}= -4\pi \epsilon \delta^{(2)}(z,\zb)g^{z\zb}g_{ab}}
Using \euler\ and \einsing\ we obtain from the first two terms in \wzf\
a singular, effectively two dimensional piece $\bar W_2$ whose
expression is:
\eqn\singt{\bar W_{2}^{A}= 16 \pi\epsilon a_4 \int d^4 x
\delta^{(2)}(z,\zb) \sqrt{\det(g_{ab})}
[g^{ab}\partial_{a}\phi\partial_{b}\phi
+{\textstyle{1\over2}}\phi\epsilon^{ab}\epsilon^{cd} R_{abcd}]}
We remark that in \singt\ the two dimensional $\epsilon$ -symbols
contain one inverse power of $\sqrt{\det(g_{ab})}$.

The second term in \singt\  has an invariant meaning following from
the fact that $g_{ab}$ and  $R_{abcd}$  at $z=\zb=0$ are the induced
metric and  the pull back of the Riemann tensor, respectively. Then
using the Gauss-Codazzi relation:
\eqn\gaco{R^{(2)}_{abcd}=\partial_{a}X^{i}\partial_{b}X^{j}\partial_{c}X^{k}\partial_{d}X^{l}
R_{ijkl}-g_{ij}(K_{ac}^{i}K_{bd}^{j}-K_{ad}^{i}K_{bc}^{j})}
where $R^{(2)} $ is calculated with the induced metric and
$K_{ab}^{i}$ is the second fundamental form, we  can rewrite
$\bar W_2$ as:
\eqn\songtt{\bar W_{2}^{A}=16 \pi\epsilon a_4 \int
d^{4}x\delta^{(2)}(z,\zb) \sqrt{\det(g_{ab})}[\phi R^{(2)} +
g^{ab}\partial_{a}\phi \partial_{b} \phi +\phi \Delta ]}
where $\Delta$ is the contribution of the second fundamental form given by:
\eqn\secf{\Delta=
g_{ij}[\tr(K^{i})\tr(K^{j})-\tr(K^{i}K^{j})]}
the traces in \secf\ being taken with the induced metric.

The first two terms in \songtt\ are the same  as in the usual two
dimensional Wess-Zumino action \wzt\ written in terms of the two
dimensional induced metric. Therefore they will produce the standard
Polyakov anomaly $\sigma \sqrt{\det(g_{ab})}R^{(2)}$. The additional
term has, however, a nonvanishing Weyl variation following from the
variation of the second fundamental form \deltaK:
\eqn\varsec{\delta_{\sigma}\Delta=-2\s\Delta -2 \partial_{i}\sigma\tr(K^{i})}
%
%
%
It follows that the total Weyl variation of $\bar W_2$ will be:
\eqn\weyvar{\delta_{\sigma} \bar W_{2}^{A}=16 \pi\epsilon a_4 \int
d^{4}x\, \delta^{(2)}(z,\zb)\sqrt{\det(g_{ab})}[\sigma (R^{(2)}
+\Delta) - 2\phi \partial_{i}\sigma \tr(K^{i})]}
We see therefore besides an addition $\Delta$ to the Polyakov
anomaly a signal for nonlocality in the appearance of a term which
still contains $\phi$ after the Weyl variation is taken. Indeed, if we
eliminate $\phi $ through the equation of motion we obtain an
expression for $\bar W_{2}$ proportional to:
\eqn\nonloc{ \int d^4 x\, \delta^{(2)}(z,\zb) \sqrt {\det(g_{ab})}
[R^{(2)} + \Delta]{1 \over \boxx^{(2)}}[R^{(2)} + \Delta]}
The Weyl variation of \nonloc\ contributed by the $\Delta$ terms
does not cancel $\boxx^{(2)}$ in the denominator and it remains
nonlocal. This is  in contradiction with the analytic structure of a
DNG contribution to the holographic action as we discussed it in
Section 3.2.

We can try to make the ``splitting'' into the two dimensional and four
dimensional contributions after the Weyl variation was taken, i.e.
directly in the expression
\eqn\eul{\delta_{\sigma} W_{4}= a_{4}\int d^{4}x
\sqrt{\det(g)}\sigma E_{4}}
From the expressions already used above we obtain:
\eqn\eulsplit{\delta_{\sigma} W_{4}= a_{4}\int d^{4}x \sqrt{\det(g)}
E_{4}^{\rm (reg)} + 16 \pi \epsilon a_4 \int d^{4}x\, \delta^{(2)}(z,\zb)
\sqrt{\det(g_{ab})} \sigma [R^{(2)} + \Delta]}
where $E_{4}^{\rm(reg)}$ contains the contribution of the curvature
components not having the $\delta$ functions.
The second contribution is now by definition local. However if an
anomaly is originating from an effective action it should obey the
Wess-Zumino condition following from the (abelian) algebra of Weyl
transformations:
\eqn\wzcond{(\delta _{2}\delta _{1}-\delta_{1}\delta_{2})W=0}
where $\delta_{i}$ is a short hand for  a variation with Weyl
parameter $\sigma_{i}$ for $i=1,2$.

Using \varsec\
we can verify directly that the second term in \eulsplit\ does not
satisfy the Wess-Zumino condition so also this second way of
``splitting'' is not tenable.
We remark that what prevents a consistent splitting is the presence of $\Delta$,
related to a nonvanishing second fundamental form.

It is instructive to check under which conditions the four
dimensional Euler anomaly obeys the Wess-Zumino condition and why after the
splitting the condition is not fulfilled anymore.
The Weyl variation of the Euler density is given by:
\eqn\vareul {\delta_{\sigma}\sqrt{\det(g)} E_{4}=-8 \sqrt{\det(g)}
G^{ij} \nabla_{i}\nabla_{j}\sigma }
where $G^{ij}$ is the Einstein tensor defined in \eins.
Using \vareul\ in \eul\ in order to calculate the double variation
we obtain:
\eqn\wzcon{(\delta _{2}\delta_{1}-\delta_{1}\delta_{2})W=-8 a_4 \int
d^{4}x\sqrt{\det(g)}
G^{ij}[\sigma_{1}\nabla_{i}\nabla_{j}\sigma_{2}-\sigma_{2}
\nabla_{i}\nabla_{j}\sigma_{1}]}
which is $0$ after an integration by parts, provided:
\eqn\bianchi{\nabla_{i} G^{ij} =0}
The Bianchi identity \bianchi\ is satisfied automatically for any
metric. The metric we use has, however,  a singular component \sing\
which produces Einstein tensor components which contain $\delta$-
functions \einsing\ and therefore the way \bianchi\ is satisfied is
rather special. The $j=z$ component of the Bianchi identity
\bianchi\ will contain also terms with $\delta$ functions. Their
cancellation requires:
\eqn\bsing{g^{z\zb}\partial_{\zb} G_{z z} -g^{ab} \Gamma^{c}_{az}G _{bc}=0 }
where the most  singular contribution to $G_{bc}$ is given
by \einsing.

The component in the first term following from \sing\ is:
\eqn\tsin{\tilde G_{zz}=- {1 \over 2}\partial_{z}g_{ab}
\partial_{z}[\epsilon g_{z\zb} \log(z\zb)]}
After using \tsin\ and \einsing\ in \bsing\ the identity is indeed
satisfied however this required a term with an explicit $\delta$-
function \einsing\ canceling against \tsin\ which produced a
$\delta$-function only after being acted upon by a derivative. In the
``splitting'' process \einsing\  is included in the two dimensional
piece while \tsin\ in the four dimensional one and the Wess-Zumino
condition which required \bianchi\ is not anymore obeyed. A similar
argument based on the inspection of \wzf\ shows that the vanishing
of the term containing the $\phi$ field in the Weyl variation, the
necessary condition for the correct analyticity, is controlled by
the same Bianchi identity \bianchi. When the second fundamental form
vanishes both  $\Gamma^{c}_{ab}$ and $\tilde G_{zz}$ in \bsing\
vanish separately and the Bianchi identity is satisfied trivially.

The mechanism  discussed above  which prevents a consistent
``splitting'' for the universal piece of the effective action
responsible for the Euler trace anomaly could be rather general:
the correct analyticity of the effective action probably requires
various Bianchi identities which mix singular components of the
curvatures (included in the two dimensional piece) with components
which become singular only after the application of derivatives.

We discuss now the second trace anomaly, the term in \wzf\ with
coefficient $c_{4}$. The four dimensional Weyl tensor is given by:
\eqn\weyt{C_{ijkl}=R_{ijkl}-{1 \over2}[g_{ik} R_{jl} + g_{jl} R_{ik}
-g_{jk} R_{il} - g_{il} R_{jk}] + {1 \over 6}[g_{ik} g_{jl}-g_{jk} g_{il}]R}
The singular components of $C_{ijkl} $ are:
\eqn\wsin{\eqalign{
\bar C_{z\zb z\zb}&=-{{4 \pi\epsilon} \over 3}g_{z\zb}
\delta^{(2)}(z,\zb)\qquad\qquad \bar C_{za\zb b}=-{{2 \pi\epsilon} \over3}
g_{ab}\delta^{(2)}(z,\zb)\cr
&C_{abcd}= {{4 \pi\epsilon }\over3}
[g_{ac}g_{bd}-g_{bc}g_{ad}]g^{z\zb}\delta^{(2)}(z,\zb)}}
Using the tracelessness of the Weyl tensor all the regular
components can be expressed in terms of the pullback:
\eqn\comp{g^{ab} C_{a z b \zb}=-\textstyle{1\over2}g_{z\zb}{g^{ab}{g^{cd} C_{acbd}}}
\qquad C_{z\zb z \zb}=-\textstyle{1\over2}g_{z\zb}^2{{g^{ac}g^{bd} C_{abcd}}} }
Finally, using \wsin\ and \comp\ in \wzf\ we obtain for the last term
the effectively two dimensional expression:
\eqn\typeb{\bar W_{2}^{B}= 16\pi\epsilon c_{4}\int d^{4}x\,
\delta^{(2)}(z,\zb) \sqrt{\det(g_{ab})}\,\phi\, g^{ac}g^{bd} C_{abcd}}
Now a Weyl transformation which shifts $\phi $ transforms the Weyl
tensor homogeneously and therefore the $\phi$ field is not present in
the anomaly indicating that the ``splitting'' is consistent for this
term in the effective action. Indeed, the anomaly is obtained by
replacing $\phi$ in \typeb\ by the Weyl transformation parameter
$\sigma(x^k)$ and the Wess-Zumino condition is satisfied
trivially. This is probably related to the fact that type B trace
anomalies have a trivial descent, i.e. the consistency conditions do
not require Bianchi identities. Equation \typeb\ has a consistent
two dimensional interpretation the invariant form of the anomaly
following from \typeb\ having the form:
\eqn\gwb{\delta_{\sigma} W_{2}= ct.\int d^{2}\tau \sqrt{\det(h)}
\sigma\, h^{ac}h^{bd}\, C_{ijkl}\partial_{a} X^{i}\partial_{b} X^{j}
\partial_{c} X^{k}\partial_d X^{l}}
where $h_{ab}$ is the induced metric.
The holographic representation should  produce therefore a
Graham-Witten anomaly of the form \gwb.

In the special case when $a_4=0$ in order to integrate out the type
B anomaly we need to add a Weyl invariant term to the Wess-Zumino
action as discussed at the beginning of the section.

In summary, the analytic structure of the type A (Euler)  Weyl anomaly
term in the effective action of the CFT representing the EE is
different than the one of its supposed holographic representation.
This was obtained under two working assumptions, i.e. that the
effective action could be used also for singular metrics and that a
first order expansion in $\epsilon$ is safe.

\newsec{The  back reaction and the Graham-Witten anomalies}

The holographic realization as used in Section 3 involved smooth
bulk metrics. Since in the CFT the singularity of the metric as
reflected in equation \curv\ played an essential role we would like
to examine if in the holographic realization such singular metrics
could appear and if they may have an influence on the discrepancy
discussed in the previous section. Of course the boundary value of
the metric $ g^{\sss(0)}$ is smooth but the solution in the bulk can
acquire singular components if the back reaction of the DNG
component of the action is taken into account.

In section 3  we have treated the dynamics of the bulk independently
producing a solution $ g_{ij}(x ,\rho)$. The embedded surface
evolved in this bulk background following the dynamics prescribed
by the DNG action. In this section we will take back reaction of the
hypersurface on the bulk into account, solve the coupled equations
of motion and evaluate the ${\cal O}(\rho^{-1})$ term of the on-shell
action. According to \HS\ this computes the Weyl anomaly, in addition
to the contribution coming from the DNG piece.

The total action is
\eqn\totalaction{
S=\int_{\cal M} d\rho\, d^d\!x\sqrt{G}(R-2\Lambda)+\int_\Sigma d\tau\, d^k\!y\,\sqrt{h}}
The equation of motion for the metric can be cast in the form
\eqn\eomtotal{
\sqrt{G}\left({1\over2}(R-2\Lambda)G^{\mu\nu}-R^{\mu\nu}\right)
=\Delta^{\mu\nu}}
with
\eqn\defelta{
\Delta^{\mu\nu}=-{1\over2}\int_\Sigma\sqrt{h}h^{\mu\nu}
\delta^{(d)}(x-X(\tau))\,\delta(\rho-\tau)}
Inserting its trace into the action results in
\eqn\onshelaction{
S=2d\int_{{\cal M}} \sqrt{G}+{d-k-2\over d-1}\int_\Sigma\sqrt{h}}
where the second term vanishes if ${\rm codim}(\Sigma)=2$, which is the
case of interest where $d=4$ and $k=2$. However, the DNG piece of the action
will feed back, through the equations of motion, into the coefficients
$g^{\sss(n)}_{ij}$.

Using the FG expansion \FGexpansion\ one finds the following expression for
the on-shell action at ${\cal O}(\rho^{-1})$ \HS\
\eqn\rhoexp{
{1\over 4}\sqrt{\det g^{\sss(0)}}\Big(\tr\gt-{1\over 2}\tr(\go^2)+{1\over4}(\tr\go)^2\Big)}
For $\go$ one finds, by solving the $(ij)$-component of \eomtotal\ at leading
non-trivial order in its $\rho$-expansion
\eqn\gonenew{ \go_{ij}={1\over2}\Big(\Rz_{ij}-{1\over6}\Rz
\gz_{ij}\Big) + \delta\!\!\go_{ij}}
where
\eqn\corrg{ \delta\!\go_{ij}=-{1\over4\sqrt{\gz}}
\int d^2 y\sqrt{\hz}\Big(\hz_{ij}-{2\over3}\gz_{ij}\Big)\delta^{(4)}(x-\Xz(y))}
and
\eqn\defh {\hz_{ij}=\gz_{\!ik}\gz_{\!jl}\hz^{ab}\,\partial_{a}\!\Xz^{k}\,\partial_{b}\!\Xz^{l}}
Note that $\delta g^{\sss(1)}$ is Weyl invariant and $g^{\sss(1)}$
is no longer universal. Consistency
with PBH and dimensional arguments restrict the most general
nonuniversal addition to $g^{\sss(1)}$, which would result for general bulk
and hypersurface action to the above form, but with arbitrary
coefficients for $h^{\sss(0)}$ and $g^{\sss(0)}$  in \corrg.

To find $\tr(g^{\sss(2)})$ it suffices to solve the $(\rho\rho)$-component of
\eomtotal\ at lowest non-trivial order:
\eqn\gtwonew{
\tr(\gt)={1\over4}\tr(\go\,^2)-{1\over32}{1\over\sqrt{\gz}}\int
d^2 y\sqrt{\hz}K^i K^j\gz_{ij}\delta^{(4)}(x-\Xz)}
The expressions \gonenew\ and \gtwonew\ represent singular
contributions to the bulk metric solution. Using the singular
contributions to  linear order in the $\delta$-function we will find
the contributions to the Graham-Witten anomaly. Quadratic and
higher order terms in the $\delta$-functions require a
regularization producing local counterterms which do not influence
the anomalies.

We find for \rhoexp\ for the case $d=4,k=2$
\eqn\anomalyfin{{1\over2} \sqrt{\gz}\Big((\tr\go)^2-\tr(\go\,^2)\Big)\Big|_{\rm \sss universal}
-{1\over16}\int d^2 y\sqrt{\hz}(K^i K^j\gz_{ij}-4\hz^{ab}\go_{ab})\,\delta^{(4)}(x-\Xz)}
where the universal $g^{\sss(1)}$ was given in \gone. Again there was a
crucial cancellation, related to the one observed above,
for ${\rm codim}(\Sigma)=2$.

Compare \anomalyfin\ to \GWktwo: we have shown that for the simplest
bulk and hypersurface actions, taking into account the back-reaction
leads to the same GW anomaly for the total action \totalaction.

The above result has a simple explanation which will allow us to
generalize the result for arbitrary bulk actions. The contributions
to the bulk metric  specified above once inserted in the bulk action
to linear order in the $\delta $-function produce a term localized
on the submanifold. Moreover this term has the same symmetries as
the DNG action. Therefore we can use the procedure discussed in
Section 3.2. The additional Graham-Witten anomaly is given by
an expression analogous to \GWanomaly\ the DNG integrand being
replaced by the term of the bulk action specified above. We will
need therefore just the expansion to order $\tau^2$ (order $\rho^2$
in our gauge) of the integrand in the first term in \totalaction.
Remembering that $\gt$  does not appear in the expansion we get the
following terms (the curvatures are computed with $g^{\sss(0)}$):
\eqn\term{ \tr(\go^{2})-(\tr \go)^{2} -\go_{ij}\, R^{ij}
+{1 \over 2} R\, \tr\go}
In the expression \term\ we left out terms in which derivatives act
on $g^{\sss(1)}$. We will discuss them in the general setting.

Now, using \gonenew\ in \term\ it is easy to verify that all the
terms linear in $\delta g^{\sss(1)}$ vanish without any need  to specify the
exact coefficients in $\delta g^{\sss(1)}$.
What is the reason for this vanishing? As we discussed in
Section 3.2  an expression  obtained by \GWanomaly\ satisfies
automatically the Wess-Zumino condition.
Independently of the exact form of the bulk action for dimensional reasons 
the only expressions which could appear in \term\
linear in $\delta g^{\sss(1)}$ are $R_{ab}$ -- the pullback of the Ricci
curvature or $R$ -- the bulk scalar curvature restricted to the
submanifold. Indeed  they do appear in individual terms in
\term. However once they are multiplied with the Weyl parameter
$\sigma $  it is easy to check that they do not fulfill the Wess-Zumino condition
and therefore they must cancel in the full expression.

Finally we return to the derivative terms left out above. Again
by a dimensional argument verified explicitly for the aforementioned
terms these contributions have the form $\boxx \sigma$
or $K^{i}\partial_{i}\sigma$, restricted to the submanifold.
These expressions do satisfy the Wess-Zumino
condition but they are cohomologically trivial being the
variations of local expressions  as we discussed in Section 3.2.

In conclusion, for an arbitrary bulk action in $d=5$ and an arbitrary  
three dimensional
DNG action the Graham-Witten anomalies remain unchanged after the
back reaction on the bulk metric is included. This is a
consequence of the fact that the Graham-Witten anomalies classified
in section 3.2 cannot originate from the $\go$ back reaction term
the only one available in $d=5$.\foot{Exactly the same argument leads 
to the identical conclusion for 
arbitrary codimension. This is so because for $k=2$ for dimensional reasons 
only $\delta\g^{\sss(1)}$ of the form \corrg, but with arbitrary 
coefficients for $h^{\sss(0)}$ and $\g^{\sss(0)}$, can contribute. 
We expect that one can also relax the condition $k=2$, in other words, 
that the back reaction never changes the GW anomalies. This was 
observed and explained in \BFS\ for conventional matter 
couplings to bulk gravity. We thank K. Skenderis for illuminating email 
exchange on this issue.} 

\newsec{Discussion}

The holographic representation of the entanglement entropy by adding
to the bulk action a DNG type action is problematic for the reasons
discussed in the previous sections. The analytic structure of
certain terms in the effective action in the CFT is different than
the one  obtained through the aforementioned  holographic mapping.
The difference in the analytic structure leads to quantitative
discrepancies even in the simplest case. We illustrate this fact by
a calculation of a certain term in the entanglement entropy when our
CFT is formulated in a flat metric background and the   two regions
in space are the exterior and interior of a sphere of radius $\bar
r$.

The entanglement entropy depends on $\bar r$ through terms
containing an ultraviolet cut off. In addition there is the
possibility of a universal logarithmic dependence. To put in
evidence this term one considers  constants appearing in:
\eqn\eel{\bar r {d \over {d \bar r}}S }

In this form it is clear  that \eel\ is related to the trace anomaly,
a change in scale of $\bar r$  being produced by a joint constant Weyl
rescaling and a rescaling diffeomorphism.  A constant Weyl
transformation produces a non zero result if the anomaly is type B
and likewise for the type A Euler density  provided that the manifold has
a nonzero Euler characteristic, which is the case for a sphere.

On the CFT side the calculation is straightforward: besides the
bulk part which is identically $0$ for a flat metric involving to
order $\epsilon $  only $R{...}$ and $\tilde R{...}$  there are the
singular contributions \eulsplit\ and \typeb:
\eqn\con{\bar r { d \over {dr}}S=16\pi\int d^{4}x\, \delta^{(2)}(z,\zb)
\sqrt{\det(g_{ab})}[a_4 (R^{(2)}+\Delta)+c_4
g^{ac}g^{bd} C_{abcd}]}
Now $C_{abcd} $ vanishes for a flat metric  and so does the first
term in \con\ since it really represents $R_{abcd}$.  The vanishing
of $R_{abcd} $ explains the relation:
\eqn\triv{ R^{(2)} = -\Delta}
where $\Delta$ is given by \secf\ which can be verified directly for
a sphere embedded in a flat metric.
As a result the constant which could have appeared in
\eel\ vanishes.

On the holographic side the contribution to \eel\ can come from one
of the Graham-Witten anomalies we studied in Section 3. The
expressions for the two type B anomalies, the pull back of $C_{ijkl}
$ and $ g_{ij}[ {1 \over 2}\tr(K^{i}) \tr(K^{j}) -\tr({K^i}K^{j})]$
vanish for the sphere as can be checked explicitly. On the other
hand the type A anomaly $R^{(2)}$ does not vanish and integrates to
the Euler number of the sphere. As we discussed in Section 3 this
anomaly does not vanish for the DNG action and it is there even for
very general forms of the three dimensional action having the
symmetries of the DNG action.  By  ``fine tuning''  the additional
terms the coefficient of $R^{(2)} $ can presumably be made to vanish but the
presence of these terms in a systematic large $N$ expansion is not
justified. We have therefore a clear contradiction between the
field theoretical calculation and its proposed holographic
representation.

Of course the DNG action can represent holographically other,
``generalized  Wilson loop type'' observables  in the CFT as
discussed in \HStwo,\Gustavsson,\Asnin. As shown in the
aforementioned references these observables have Graham-Witten
anomalies completely compatible with the ones produced by the
holographic representation.

We are faced therefore with the problem of producing a
holographic  representation of the entanglement entropy which is
compatible with the field theoretical constraints.
An obvious guess would be simply a bulk gravitational action
whose classical solution matches at the boundary the singular metric
studied in Section 2.

As far as the terms in the action producing the trace anomalies are
concerned this proposal seems to be valid, though in a rather
tautological way: as we discussed in Section 3 the calculation of
the trace anomalies does not require the solution of the equations
of motion but just the evaluation of some boundary terms for the
boundary metric. Therefore the holographic calculation is bound to
reproduce the results of Section 4.

For other terms of the action the solution of the equations of
motion with singular boundary conditions would be needed and it is far from obvious
that such a calculation can be controlled.

A related question is the appearance and interpretation of non
universal contributions in the effective action of the CFT
representing the EE. As a concrete example we consider again the EE
for a sphere embedded in a flat metric. On general grounds \EE\ one
expects for the EE a leading dependence proportional to $\bar r^2$
multiplied by an appropriate scale. Are such terms obtainable in the
CFT from Weyl invariant contributions to the effective action which
however being more singular than the terms we considered require an
additional regularization for the singular metric or are simply, non
universal subtractions (boundary terms) like in the  holographic
representation?

These and other
questions related to  the holographic representation of the
EE are presently under study.

{\bf Acknowledgments} Very useful discussions with M. Ba\~nados, N.
Boulanger, D. Giulini, D. Kutasov, H. Neuberger, S.
Shankaranarayanan and H. Shimada are gratefully acknowledged.

\vfill\eject

\noindent {\bf Appendix: Derivation of the universal type A
anomaly coefficient}

In \ISTY\ it was shown that for any gravitational action \bulkaction\ with
has $AdS_{2n+1}$ as a solution to the equations of motion, the coefficient $a$ of the
unique type A anomaly of dual CFT is
$$
a_n={b_0\over 2^{2n}(n!)^2}\eqno(A.1)
$$
where $b_0=f(AdS)$. In \ISTY\ this was derived by looking at a
conformally flat metric $\gz$ and solving the PBH-transformation
equation for $b_n$. Here we will present an alternative derivation
which uses the ideas of \Boulanger. There it was observed that while
the type B Weyl anomalies have a trivial descent, the unique (in any
even dimensions) type A anomaly has a non-trivial descent.\foot{The
descent of cohomologically trivial contributions stops  after the
second step.} These features might, in fact, serve as the defining
distinction between the two classes of anomalies, which can also be
applied to the hypersurface anomalies discussed in sect. 3.

We begin with a review of the results of \Boulanger.
Define
$$
\eqalignno{
{\cal O}^{j_1\dots j_p}_{1 2 \dots p+1}={4^p\, n!\over 2^n (n-p)!}
\sqrt{g}g_{i_1 k_1}\dots g_{i_p k_p}&
\epsilon^{i_1 j_1\dots i_n j_n}\epsilon^{k_1 l_1\dots k_n l_n}&(A.2)\cr
&\times\,R_{i_{p+1} j_{p+1} k_{p+1}l_{p+1}}\cdots R_{i_n j_n k_n l_n}
\s_{[1}\p_{l_1}\s_2\dots \p_{l_p}\s_{p+1]}}
$$
where the antisymmetrization is over the indices of $\s$.
In particular
$$
{\cal O}_1=\s_1 \sqrt{g}\,E_{2n}\eqno(A.3)
$$
with
$$
E_{2n}={1\over 2^n}\epsilon^{i_1 j_1\dots i_n j_n}\epsilon^{k_1 l_1\dots k_n l_n}
R_{i_1 j_1 k_1 l_1}\cdots R_{i_n j_n k_n l_n}\eqno(A.4)
$$
the $d$-dimensional Euler density. The normalization is such that
$E_{2n}=R^n+\dots$. ${\cal O}_1$ is at the top of the descent which is
$$
\delta^{\phantom{j_1}}_{[{p+1}}{\cal O}^{j_1\dots j_{p-1}}_{1\dots p]}
=\p^{\phantom{j_1}}_{p}{\cal O}^{j_1\dots j_{p}}_{1\dots p+1}\eqno(A.5)
$$
and
$$
{\cal O}^{j_1\dots j_n}_{12\dots n+1}
=2^n (n!)^2\sqrt{g}\,\s_{[1}\nabla^{j_1}\s_2\dots\nabla^{j_n}\s_{n+1]}\eqno(A.6)
$$
is at the bottom.
In deriving (A.5)
we need the Weyl variation of the Riemann tensor
$$
\delta R_{ijkl}=2\s R_{ijkl}+g_{ik}\nabla_j\nabla_l\s+g_{jl}\nabla_i\nabla_k\s
-g_{il}\nabla_j\nabla_k\s-g_{jk}\nabla_i\nabla_l\s\eqno(A.7)
$$

The holographic version of the descent starts with the
$d+1$ dimensional `CS-form' ${\cal O}=\sqrt{G}f(R)$.
Under PBH
$$
\delta_1{\cal O}=\p_\mu(\xi_1^\mu{\cal O})\equiv \p_\mu{\cal O}_1^\mu\eqno(A.8)
$$
If we define
$$
{\cal O}^{\mu_1\dots \mu_p}_{1\dots p}=\xi^{\mu_1}_{[1}\cdots\xi^{\mu_p}_{p]}{\cal O}\eqno(A.9)
$$
we can show, using the group property \groupproperty
$$
\delta_{p+1}{\cal O}^{\mu_1\dots \mu_p}_{1\dots p}
=\p_{\mu_{p+1}}{\cal O}^{\mu_1\dots\mu_{p+1}}_{1\dots p+1}\eqno(A.10)
$$
Using \PBHaction\ for the $\rho$-expansion of ${\cal O}$ and
$\xi^\rho=2\s\rho\,,\xi^i={1\over2}\rho g_{(0)}^{ij}\p_j\s+{\cal O}(\rho^2)$
we find
$$
{\cal O}^{\rho j_1\dots j_n}_{1\dots n+1}={1\over 2^n}\sqrt{g}\, b_0\,
\s_{[1}\nabla^{j_1}\s_2\cdots\nabla \s_{n+1]}^{j_n}\eqno(A.11)
$$
Comparing this with (A.6) we conclude that the holographic type A Weyl anomaly
in $d=2n$ dimensions is $a_n E_{2n}$ with $a_n$ as in (A.1).

\listrefs

\bye